\documentclass[aps,pre,twocolumn,nofootinbib,groupedaddress,superscriptaddress,showpacs]{revtex4-1}

\usepackage{graphicx}
\usepackage{color}
\usepackage{amsmath}
\usepackage{amsfonts}
\usepackage{amssymb}
\usepackage{dcolumn}
\usepackage{hyperref}
\usepackage{enumerate}
\usepackage{bbold}
\usepackage{cancel}
\usepackage{mathtools}
\begin{document}

\title{A Random-Line-Graph Approach to Overlapping Line Segments}
\author{Lucas B\"ottcher}
\affiliation{Computational Medicine, UCLA, 90095-1766, Los Angeles, United States}
\affiliation{Institute for Theoretical Physics, ETH Zurich, 8093, Zurich, Switzerland}
\affiliation{Center of Economic Research, ETH Zurich, 8092, Zurich}
\email{lucasb@ethz.ch}
\date{\today}
\begin{abstract}
We study graphs that are formed by independently-positioned needles (i.e., line segments) in the unit square. To mathematically characterize the graph structure, we derive the probability that two line segments intersect and determine related quantities such as the distribution of intersections, given a certain number of line segments $N$. We interpret intersections between line segments as nodes and connections between them as edges in a spatial network that we refer to as random-line graph (RLG). Using methods from the study of random-geometric graphs, we show that the probability of RLGs to be connected undergoes a sharp transition if the number of lines exceeds a threshold $N^*$.
\end{abstract}
\maketitle
\section{Introduction}
We study the properties of graphs that result from intersecting needles (i.e., line segments), which are independently-positioned in the unit square. To do so, we interpret intersections as nodes in a spatial network~\cite{barthelemy2018complex} and connections between them as edges. We refer to these networks as random-line graphs (RLGs). The structure of RLGs is similar to that of assemblies of overlapping rods, which have been studied in the context of continuum percolation~\cite{quintanilla1996clustering,mertens2012continuum,schilling2015percolation,xu2016continuum}. Another related process is random-sequential adsorption (RSA)~\cite{evans1993random} where ``particles'' are randomly deposited on a surface and adsorbed if they do not overlap any previously adsorbed particle~\cite{vigil1990kinetics,ziff1990kinetics,manna1991random,vsvrakic1991kinetics,bonnier1994adsorption}.

Random-line graphs are constructed in a similar way as two-dimensional random-geometric graphs (RGGs)~\cite{dall2002random,penrose2003random}, whose nodes (i.e., points) are connected if their distance is smaller than a certain threshold. Note that other choices of connection functions are also possible~\cite{dettmann2016random}. One major difference between the two random-graph models is that nodes in an RLG result from the intersection of line segments (i.e., possible edges) whereas nodes in an RGG are initially given and edges between them are only established according to certain connection functions.

Random-geometric graphs and their modifications have found various applications as models of social interaction networks~\cite{banisch2019opinion}, wireless ad hoc and sensor networks~\cite{glauche2003continuum,coon2012impact,dettmann2016random}, and networks of neurons~\cite{penrose2003random}. Modified RGG models~\cite{nauer2019random} were also used to model granular networks that are usually constructed by interpreting particles of a granular packing as nodes and contacts between them as edges~\cite{papadopoulos2018network}. Models of granular networks are helpful to mathematically and computationally study the ability of network analysis methods to characterize physical transport characteristics, failure mechanisms, and other system-level properties of granular packings~\cite{smart2007effects,smart2008granular,berthier2019forecasting}. In general, the application of network analysis methods to granular systems may contribute to improvements in fracture control, material design, and understanding deformation effects~\cite{herrmann2014statistical,anderson2017fracture,heisser2018controlling,mungan2019networks}.
Such network-based approaches complement existing continuum~\cite{anderson2017fracture} and particle-level~\cite{herrmann2014statistical,cates1999jamming} descriptions of granular systems~\cite{jaeger1996granular,andreotti2013granular} by also considering the intermediate-scale organization and interaction of particles~\cite{papadopoulos2018network}.

Due to the broad applications of RGGs and their modifications, we expect the study of RLGs to be relevant to obtain further insights into transport and communication networks~\cite{barthelemy2018complex} and related biological systems such as fibrin networks~\cite{munster2013strain,van2014constitutive}, which are the main structural components of blood clots. The formation of fibrin networks exhibits some similarities to the construction of RLGs. During the formation process, fibrin monomers polymerize into fibers that intersect with each other and form branching points~\cite{van2014constitutive}.

To study the properties of RLGs, we first derive the probability that two line segments intersect and determine the distribution of intersections for a certain number of lines $N$. We also numerically study cluster-size distributions and derive an upper bound for the  threshold above which RLGs are connected with finite probability.
\section{Intersecting Needles and Random-Line Graphs}
\label{sec:model}
\begin{figure*}
\includegraphics[width=0.32\textwidth]{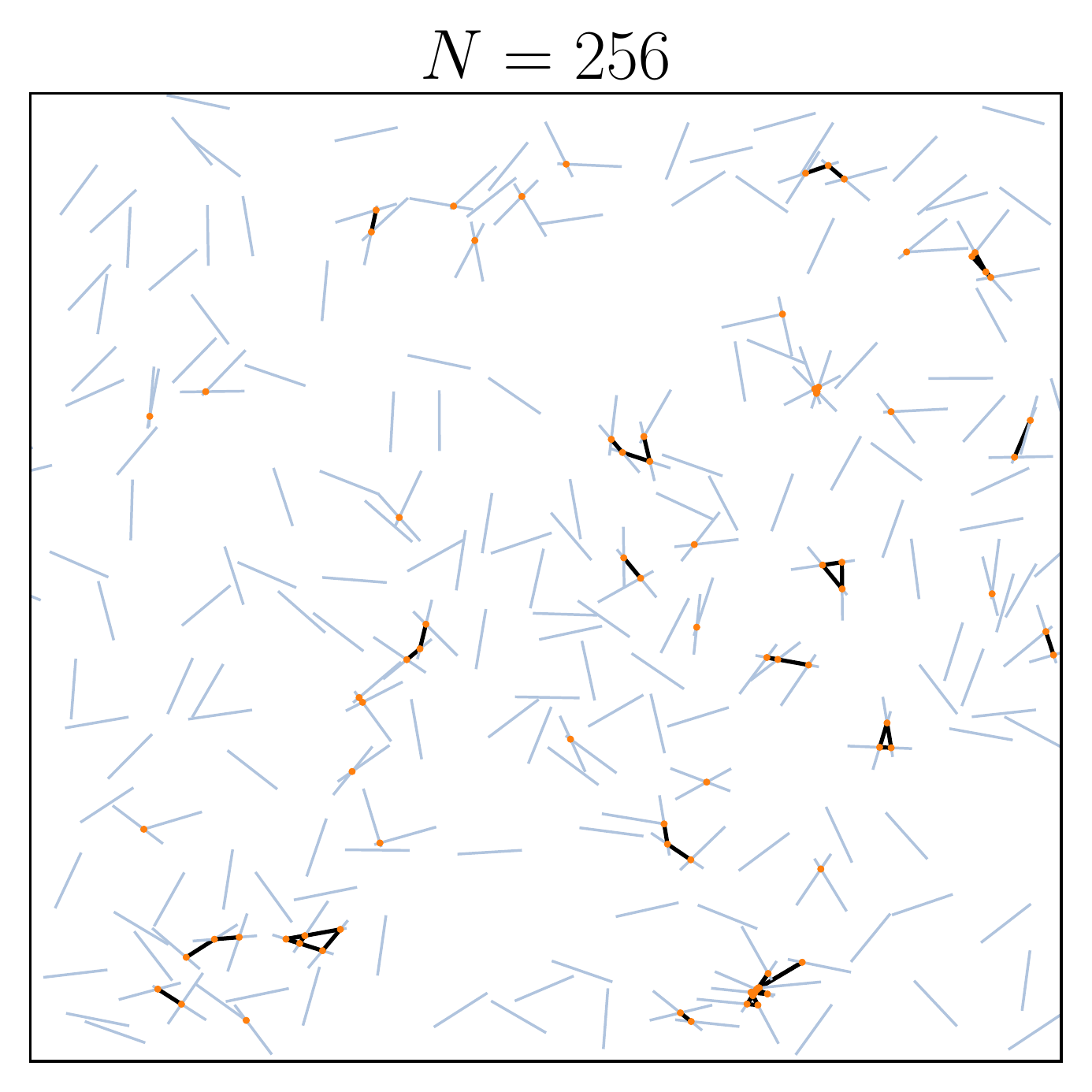}
\includegraphics[width=0.32\textwidth]{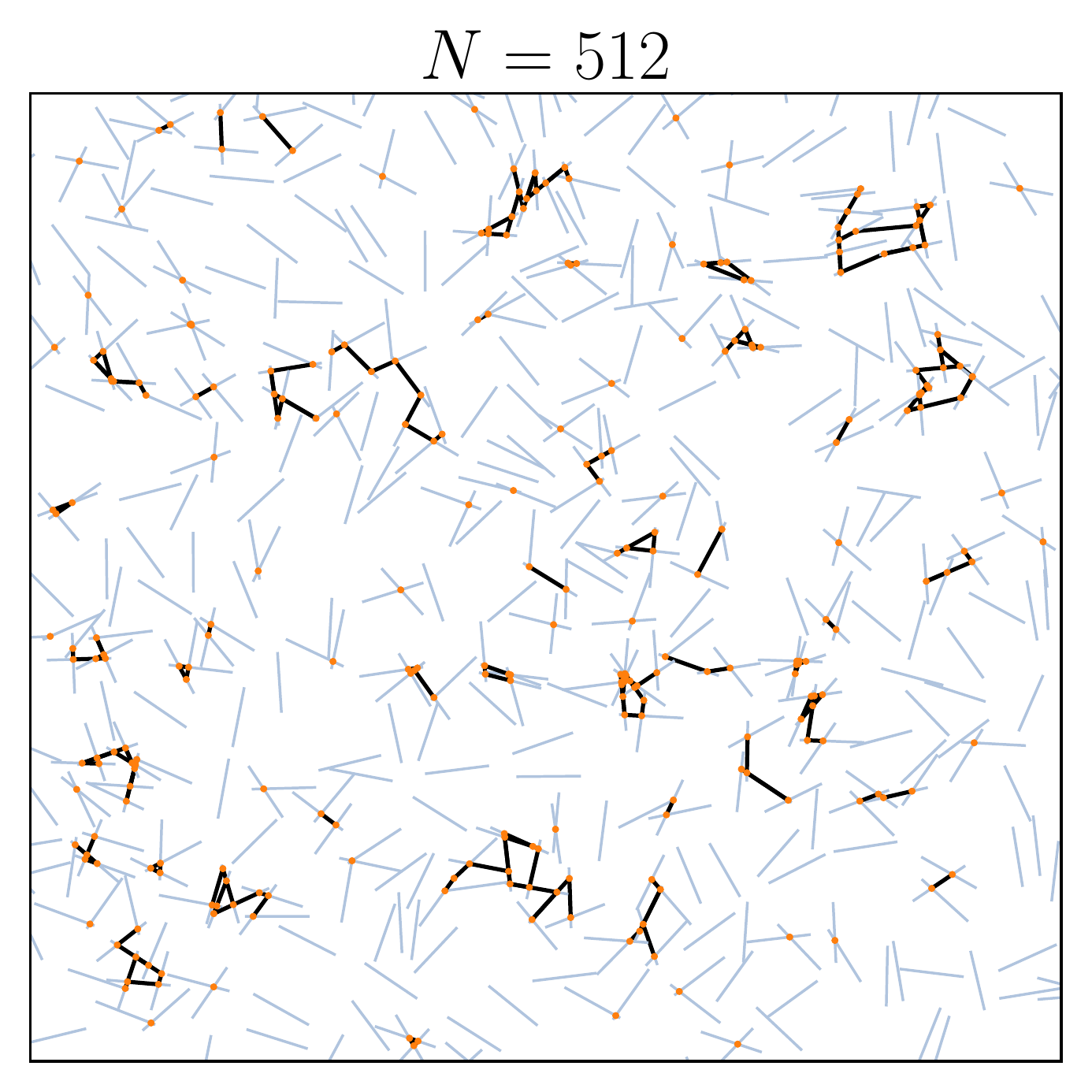}
\includegraphics[width=0.32\textwidth]{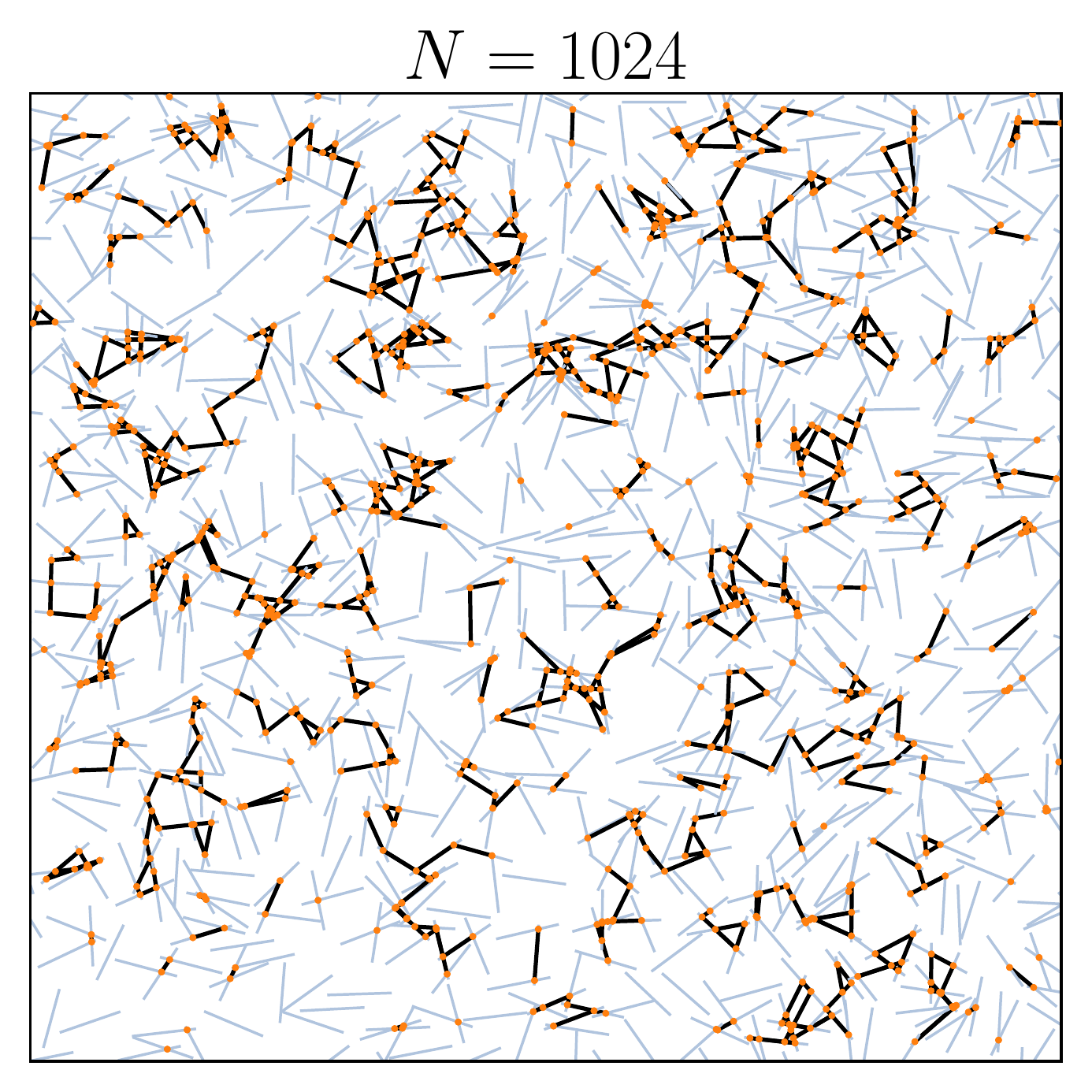}
\caption{\textbf{Examples of RLGs.} We show three realizations of RLGs that result from the intersection of 256, 512, and 1024 line segments of length $a=0.06$ (blue). Intersections of blue line segments correspond to nodes (orange dots) in an RLG. Nodes are connected by an edge (black line) if they are located on the same blue line segment. The number of nodes and edges $(n,m)$ is: $(89,69)$ for $N=256$, $(285,250)$ for $N=512$, and $(1048,1225)$ for $N=1024$.}
\label{fig:line_networks}
\end{figure*}
To construct RLGs, we independently position $N$ line segments of length $a$ in the unit square $[0,1]^2$. To do so, we draw the origin of each line $(x_1,y_1)$ from the uniform distribution (i.e., $x_1,y_1 \sim \mathcal{U}(0,1)$) and determine the corresponding endpoint according to $(x_2,y_2)=(x_1+a \cos(\varphi),y_1+a \sin(\varphi))$ with $\varphi \sim \mathcal{U}(0,2\pi)$.  We denote the graph (i.e., network) that results from the intersections of $N$ lines by $G(N,a)$ and use $n$ and $m$ to denote the corresponding number of nodes and edges. An intersection between two lines $i\in \{1,\dots,N\}$ and $j\in\{1,2,\dots,i-1,i+1,\dots,N\}$ (and vice versa) corresponds to a node with spatial coordinates $(x_{i j},y_{i j})$. The maximum number of nodes in an RLG is $N(N-1)/2$. If two nodes are connected by one of the $N$ line segments, an edge is created between them. In Fig.~\ref{fig:line_networks}, we show three realizations of RLGs for a line segment length of $a=0.06$ and $N=256$, $512$, and $1024$. Orange dots in Fig.~\ref{fig:line_networks} represent nodes and black lines the corresponding edges. For $256$ line segments (blue), only a few intersections occur and a substantial portion of nodes is isolated. If we increase the number of lines to $512$ and $1024$, we find that there are more intersections and the relative number of isolated nodes becomes smaller. We show in Sec.~\ref{sec:connected} that the probability of observing isolated nodes decreases with $N$ and eventually becomes vanishingly small. In particular, if the number of lines exceeds a threshold $N^*$ for a given length $a$, the probability of an RLG to be connected undergoes a transition that can be also observed in RGG models. This behavior is reminiscent of the emergence of wrapping clusters in continuum and line/stick percolation~\cite{li2009finite,mertens2012continuum,speidel2018topological,balister2018line}.
\section{Intersection Probability}
\begin{figure*}
\includegraphics{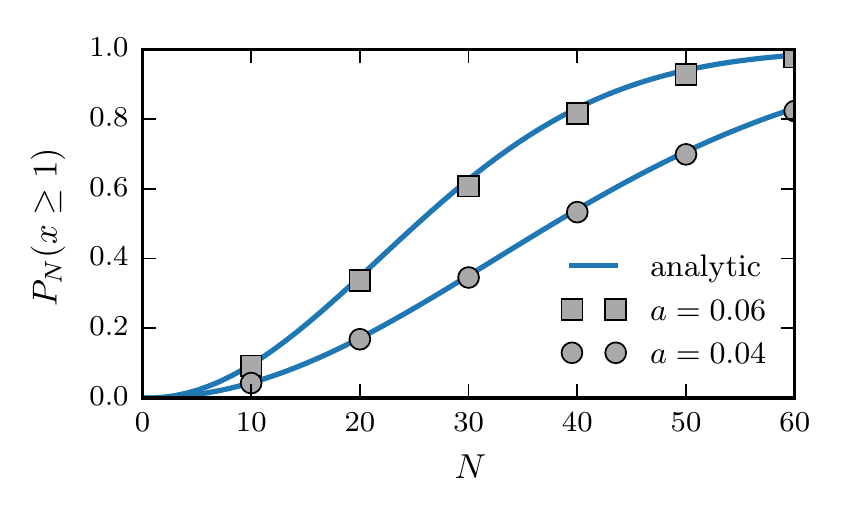}
\includegraphics{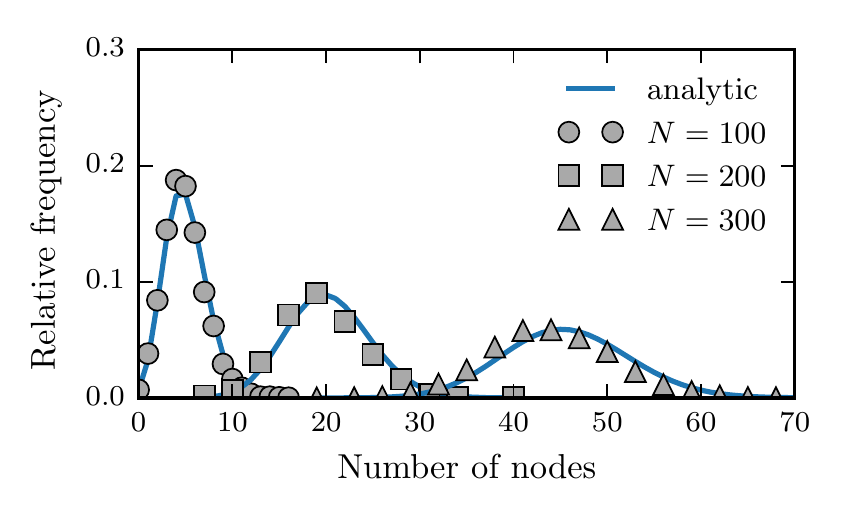}
\includegraphics{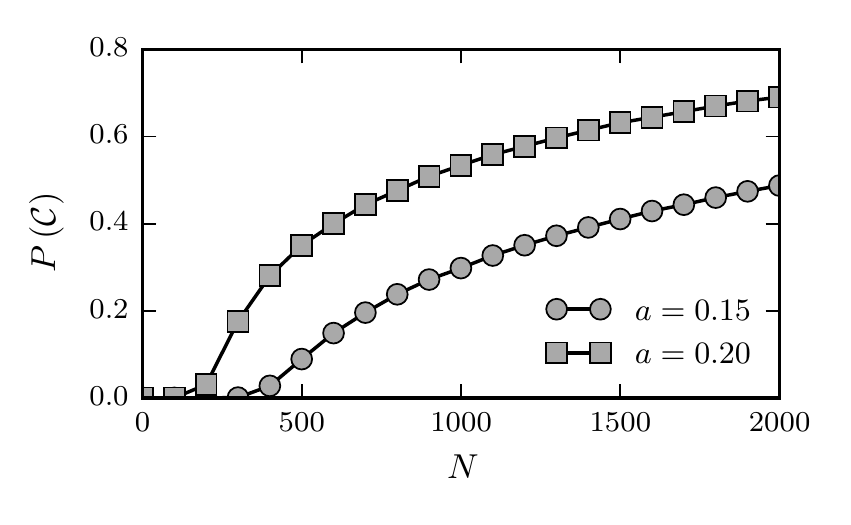}
\includegraphics{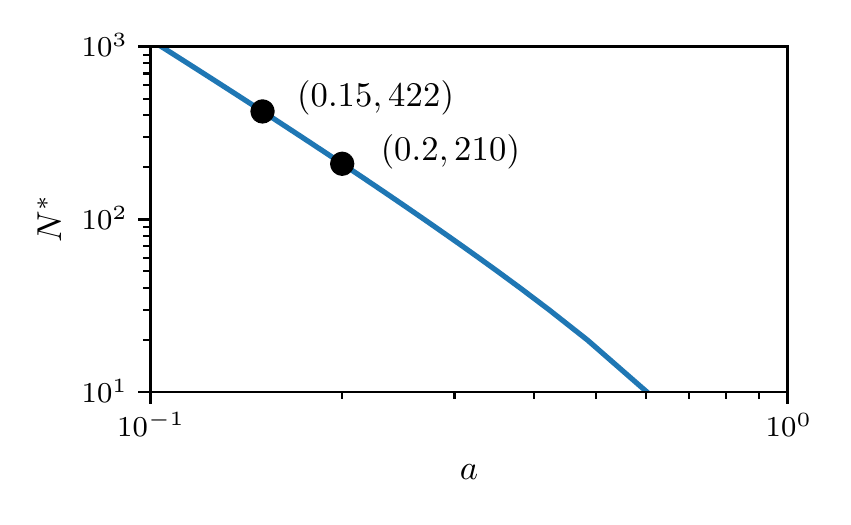}
\caption{\textbf{Intersection probability and connectivity.} In the top left panel, we show the probability $P_N(x\geq 1)$ that at least one out of $N(N-1)/2$ possible intersections occurs for line segment lengths $a=0.04$ and $a=0.06$. The blue solid line is the analytical result of Eq.~\eqref{eq:intersect_prob} and markers indicate numerical results that are based on
$10^5$ samples. For a line segment length of $a=0.04$, we show the distribution of intersections (i.e., nodes in an RLG) for different numbers of line segments $N$ in the top right panel. The data is based on $10^5$ samples and blue solid lines correspond to the Poisson approximation of Eq.~\eqref{eq:Poisson_nodes}. 
In the bottom left panel, we show the probability $P(\mathcal{C})$ that an RLG is connected as a function of the number of line segments $N$ for $a=0.15$ and $a=0.2$. We considered $2\times 10^5$ realizations of RLGs to compute $P(\mathcal{C})$. In the vicinity of $N^*$, RLGs undergo a sharp transition from being a.a.s.\ disconnected to being a.a.s.\ connected. We show the functional dependence of $N^*$ on $a$ in the bottom right panel and also mark the two line segment length thresholds that correspond to $a=0.15$ and $a=0.2$ with black dots.
}
\label{fig:numerics_analytics}
\end{figure*}
An important quantity for the mathematical characterization of RLGs is the probability $p(a)$ that two line segments of length $a$ intersect with each other. Two line segments can intersect if they are both located in a region of area $A=a^2(2+\pi)$ (see Fig.~\ref{fig:overlap_area}). That is, the origin of an intersecting line segment has to be located either in a rectangular-shaped region of area $2 a^2$ or in one of the four quarter circles of area $\pi a^2/4$ surrounding the target line segment. We derive the intersection probability for all sub-regions in App.~\ref{app:intersection_prob} and obtain
\begin{equation}
p(a)=\frac{2 a^2}{\pi}\,.
\label{eq:intersect_prob}
\end{equation}
Note that the derivation of Eq.~\eqref{eq:intersect_prob} is structurally similar to the derivation of the intersection probability in Buffon's needle problem where needles of length $a$ are dropped on a plane with infinitely long parallel strips. Given that the parallel strips are a distance $b>a$ apart, the corresponding intersection probability is $2 a/(b \pi)$~\cite{krauth2006statistical}.

We also outline in App.~\ref{app:intersection_prob} that the probability that an intersection originates from one of the quarter circles of Fig.~\ref{fig:overlap_area} is three times smaller than the intersection probability in the square-shaped regions. For the derivation of Eq.~\eqref{eq:intersect_prob}, we implicitly assume that the length $a$ of a line segment is small enough so that almost all intersections occur within the considered domain. If the domain is not the unit square, Eq.~\eqref{eq:intersect_prob} has to be normalized by the corresponding area.

We now compare the analytically obtained expression of $p(a)$ with corresponding numerical data. Let $x$ denote the number of intersections that result from independently positioning $N$ lines in the unit square. For $N=1$, the probability $P_{N=1}(x\geq 1)$ that at least one intersection occurs is zero and $P_{N=2}(x\geq 1)=p(a)$. In general, the probability that at least one out of $N(N-1)/2$ possible intersections occurs is
\begin{equation}
P_N(x\geq 1) = 1-\prod_{i=1}^N \left[1-(i-1)\,p(a)\right]\,.
\label{eq:P_N_x}
\end{equation}
We consider two line segments of lengths $a=0.04$ and $a=0.06$ as examples and compare the analytical prediction of Eq.~\eqref{eq:P_N_x} with corresponding numerical results in the top left panel of Fig.~\ref{fig:numerics_analytics}. We find that analytical and numerical results are in good agreement. For $a=0.06$, the probability $P_N(x \geq 1)$ is almost 1 for $N \geq 60$ whereas a smaller value of $a=0.04$ yields smaller probabilities $P_N(x \geq 1)$. To further characterize the number of intersections (i.e., nodes in an RLG), we compute the distribution of node numbers for different values of $N$ and $a=0.04$ (see top right panel of Fig.~\ref{fig:numerics_analytics}). We observe that the distributions broaden with increasing number of lines $N$ and their peaks shift to the right. According to the definition of RLGs in Sec.~\ref{sec:model}, nodes result from the intersection of independently positioned line segments, and so we approximate the distribution of nodes $P_N(n)$ (i.e., the probability of observing $n$ nodes/intersections in an RLG consisting of $N$ line segments) by a Poisson distribution:
\begin{equation}
P_N(n) \approx \frac{\bar{n}^n \exp\left({-\bar{n}}\right)}{n !}\,,
\label{eq:Poisson_nodes}
\end{equation}
where $\bar{n} = p(a) N(N-1)/2$ is the mean number of nodes.
A comparison between the numerically obtained data and the Poisson approximation of Eq.~\eqref{eq:Poisson_nodes} shows that the data is well-described by the approximation for $N=100$ (see top right panel of Fig.~\ref{fig:numerics_analytics}). Deviations from the analytic approximation for $N=200$ and $300$ occur due to the finite length of line segments and intersections that occur outside of the considered domain.
\section{Connectivity}
\label{sec:connected}
The examples of RLGs that we show in Fig.~\ref{fig:line_networks} suggest that the probability of observing isolated nodes becomes vanishingly small for sufficiently large numbers of line segments $N$. To mathematically describe the \emph{sharp threshold}~\footnote{According to Ref.~\cite{friedgut1996every}, a sharp transition corresponds to a ``swift'' transition in random-graph models ``\emph{from a property being very unlikely to it being very likely}''.} at which RLGs become connected, we use methods from the study of RGGs~\cite{penrose2003random,diaz2009probability} and focus on the case where all $N$ line segments that form an RLG are connected. If all line segments are connected, the corresponding RLG is also connected. However, even if an RLG is connected, there may exist a few isolated line segments. We consider the probability that all $N$ line segments are connected as an upper bound for the probability that an RLG is asymptotically almost surely (a.a.s.) connected. The following results are asymptotic as $N\rightarrow \infty$.
\begin{figure*}
\includegraphics{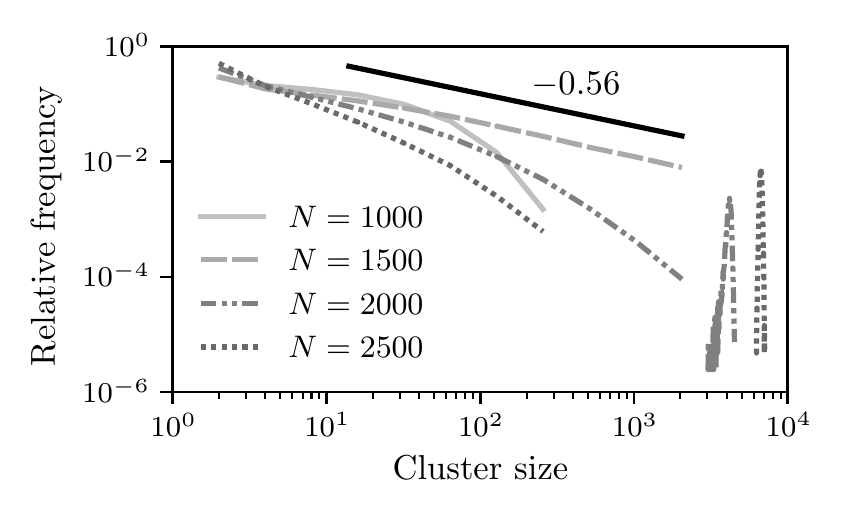}
\includegraphics{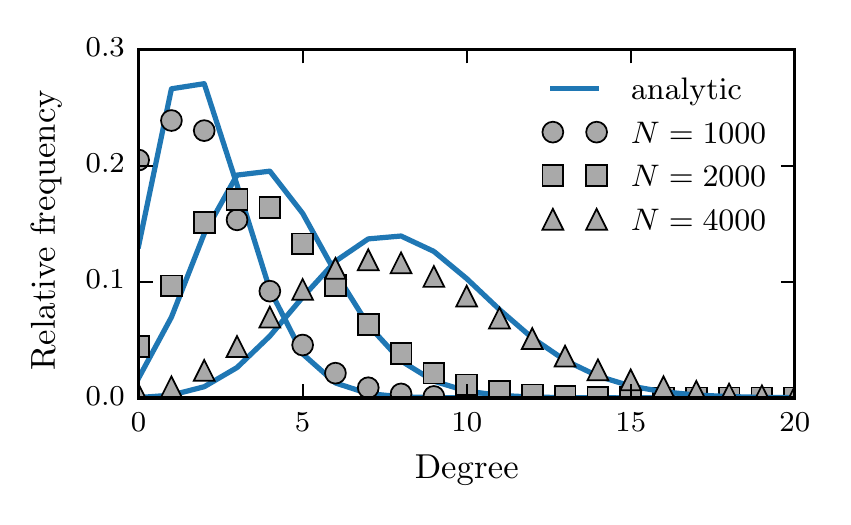}
\caption{\textbf{Cluster-size and degree distributions.} We consider RLGs with $a=0.06$ and show the relative frequency of cluster sizes for $N=1000$, $1500$, $2000$, and $2500$ in the left panel. To smoothen the data from the $2\times10^4$ realizations of RLGs, we count the cluster-size occurrences in the intervals $[2^k,2^{k+1}]$ for $k\in\{0,10\}$. For cluster sizes larger than $2048$, we use a finer binning. The black solid line is a guide to the eye with the indicated slope. We determined the slope with a least-square fit (assuming a power-law behavior) for $N=1500$ and cluster sizes that are at least $11$. In the right panel, we show the relative frequency of degrees for $N=1000$, $2000$, and $4000$. The data is based on $10^5$ realizations.}
\label{fig:cluster_size}
\end{figure*}

According to Eq.~\eqref{eq:intersect_prob}, the probability that one out of $N$ line segments is isolated is $(1-p(a))^{N-1}$. We use $X$ to denote the event that one line segment is isolated. The corresponding expectation value of isolated line segments is~\cite{diaz2009probability}
\begin{equation}
\mathbb{E}\left(X\right)=N (1-p(a))^{N-1}=N e^{-p(a) N - \mathcal{O}({p(a)}^2 N)}\,,
\end{equation}
where we used that $N \log(1+x)= N\left(x-\mathcal{O}(x^2)\right)$ in the second step. In accordance with Ref.~\cite{diaz2009probability}, we define $\mu \coloneqq N \exp\left[-p(a) N\right]$ and note that the asymptotic behavior of $\mu$ characterizes the connectivity of all $N$ line segments as in the case of RGGs. In particular, all line segments that are used to construct an RLG are a.a.s.\ connected if $\mu\rightarrow 0$ and a.a.s.~disconnected if $\mu\rightarrow \infty$. The case where $\mu = \Theta(1)$ (i.e., $\mu$ is asymptotically bounded from below and above by a constant~\footnote{In general, the notation $f(x)=\Theta(g(x))$ is used to indicate that $f(x)$ is asymptotically bounded above and below by $g(x)$.}) corresponds to a connectivity threshold. We reformulate the condition $\mu = \Theta(1)$ in terms of a threshold length~\cite{diaz2009probability}
\begin{equation}
a^*=\sqrt{\frac{\pi \log(N)\pm \mathcal{O}(1)}{2 N}}\,.
\label{eq:a_c}
\end{equation}
If $a$ approaches $a^*=\sqrt{\frac{\pi \log(N)}{2 N}}$ from below, all $N$ line segments that form a RLG become a.a.s.\ connected instead of being a.a.s.\ disconnected.

A certain threshold length $a^*$ corresponds to a threshold number of lines $N^*$ that can be obtained by numerically inverting Eq.~\eqref{eq:a_c} for a given line length $a$. We show the functional dependence of $N^*$ on $a$ in the bottom right panel of Fig.~\ref{fig:numerics_analytics}. For $a=0.15$, the threshold number of lines is $N^* \approx 422$ and for $a=0.2$ it is $N^*\approx 210$. 

We can now test the ability of the derived thresholds to describe the connectivity transition of RLGs by considering corresponding numerical data and denote by $\mathcal{C}$ the event that an RLG is connected. In the bottom left panel of Fig.~\ref{fig:numerics_analytics}, we show the probability $P(\mathcal{C})$ that an RLG is connected as a function of $N$. The shown data is based on $2\times 10^5$ realizations of RLGs. We observe that $P(\mathcal{C})$ undergoes a transition from zero to positive values in the vicinity of $N^*$. Based on the analytical and numerical insights, we use $N^*$ as an upper bound above which RLGs become a.a.s.\ connected. In the case of the considered examples, the values of $N^*$ provide a relatively accurate description for the connectivity transition point of RLGs (see bottom panels of Fig.~\ref{fig:numerics_analytics}). 

We note that $P(\mathcal{C})$ should not be confused with the wrapping probability of (stick) percolation. Above the critical line-segment density $\rho_c\approx 5.65$ of stick percolation~\cite{li2009finite,mertens2012continuum}, the wrapping probability approaches unity in an infinite system. For the line-segment length $a=0.15$ and unit-square domain that we considered in Fig.~\ref{fig:numerics_analytics} (bottom panels), the critical number of lines that corresponds to the density $\rho_c\approx 5.65$ is approximately $38$, much lower than $N^* \approx 422$. For the existence of a wrapping cluster in stick percolation, it is not necessary that all intersecting lines form a connected network. The critical number of lines in stick percolation is thus smaller than the number of lines at the corresponding connectivity threshold.
\section{Cluster-size distribution}
If the number of line segments exceeds $N^*$, an RLG mainly consists of a large connected component that contains almost all nodes and some smaller components with only a few nodes. On the other hand, for small values of $N$, we would expect many small components and only a few larger ones. To examine the cluster-size distribution for different values of $N$, we generate $2\times 10^4$ realizations of RLGs with $a=0.06$ and $N=1000$, $1500$, $2000$, and $2500$. The corresponding threshold number of lines is $N^*\approx 3569$ (see Eq.~\eqref{eq:a_c}). For each value of $N$, we determine the corresponding cluster sizes (i.e., numbers of nodes in connected components) and their relative occurrences. To smoothen the data, we count the relative occurrences of clusters in the intervals $[2^k,2^{k+1}]$ for $k\in\{0,10\}$. For cluster sizes larger than $2048$, we use a finer binning of the data. We show the resulting relative frequencies of cluster sizes for RLGs with $a=0.06$ and $N=1000$, $1500$, $2000$, and $2500$ in the left panel of Fig.~\ref{fig:cluster_size}. In the case of $N=1000$, the largest components consist of a few hundred nodes and only occur with a relative frequency of less than $10^{-2}$. However, if we increase $N$ to a value of $1500$, the relative frequencies of large clusters also increases and the corresponding distribution becomes more heavy-tailed. For even larger values of $N$, the relative frequency of large clusters increases whereas it becomes less probable to observe small clusters (see left panel of Fig.~\ref{fig:cluster_size}). This behavior indicates that the system approaches the connectivity transition point. Cluster-size distributions that are similar to the ones we observe for $N=2000$ and $2500$ have been also described in other models such as epidemic models~\cite{bottcher2016connectivity} and birth-death-immigration models~\cite{xu2018immigration}.
\section{Degree distribution}
Each intersection between two out of $N$ line segments leads to a node in an RLG and each node can have a maximum number of $N-2$ neighbors. We approximate the mean degree $\bar{k}$ of each node by assuming that the potential intersections of $N-2$ line segments with the two line segments that define a certain node are independent events and obtain $\bar{k}\approx 2 p(a) (N-2)$. Similar to the Poisson approximation of the node distribution (see Eq.~\eqref{eq:Poisson_nodes}), we approximate the degree distribution $P_N(k)$ (i.e., the relative frequency of nodes with degree $k$ in an RLG consisting of $N$ line segments) by
\begin{equation}
P_N(k)\approx \frac{\bar{k}^k \exp\left({-\bar{k}}\right)}{k !}\,.
\label{eq:Poisson_edges}
\end{equation}
In the right panel of Fig.~\ref{fig:cluster_size}, we show a comparison between the approximated degree distribution of Eq.~\eqref{eq:Poisson_edges} and corresponding numerical data. We find that the analytical approximation is able to describe characteristic features of the observed degree distributions for different numbers of line segments $N$.
\section{Discussion and Conclusion}
We studied systems that consist of $N$ line segments, which are independently distributed in the unit square. Intersections between these line segments form nodes in a planar spatial network that we refer to as RLG. We derived the probability that two line segments intersect and described intersection and cluster-size distributions. Furthermore, we used methods from the study of RGGs to determine an upper bound of the threshold above which RLGs become connected.

Our results provide new insights in spatial random-graph models and may be helpful to the study of systems, such as fibrin networks~\cite{munster2013strain,van2014constitutive}, that exhibit structural similarities to RLGs. Similar to a variant of Buffon's needle problem that considers curved needles (i.e., ``Buffon's \emph{noodle} problem''), future studies may explore variations in the shape distribution of lines that form RLGs. In addition to different line shapes, another possible direction for future research is to study how shape variations in the box sizes~\cite{estrada2015random} affect network characteristics of RLGs. Furthermore, the derived analytical expressions of the intersection probability of line segments and the corresponding connectivity threshold may be useful for the study of continuum percolation problems~\cite{mertens2012continuum,speidel2018topological}.
\acknowledgements{The author thanks Ehsan Reyhanian for outlining an alternative proof of Eq.~\eqref{eq:intersect_prob} and Mason A.~Porter for very helpful suggestions and detailed feedback. Numerical calculations were performed on the ETH Euler cluster. The author acknowledges financial support from the SNF Early Postdoc.Mobility fellowship on ``Multispecies interacting stochastic systems in biology'' and the Army Research Office (W911NF-18-1-0345).}
\clearpage

\onecolumngrid

\appendix
\renewcommand\thefigure{\thesection.\arabic{figure}}
\section{Intersection probability}
\label{app:intersection_prob}
\setcounter{figure}{0}    
\begin{figure}
\centering
\includegraphics[width=0.8\textwidth]{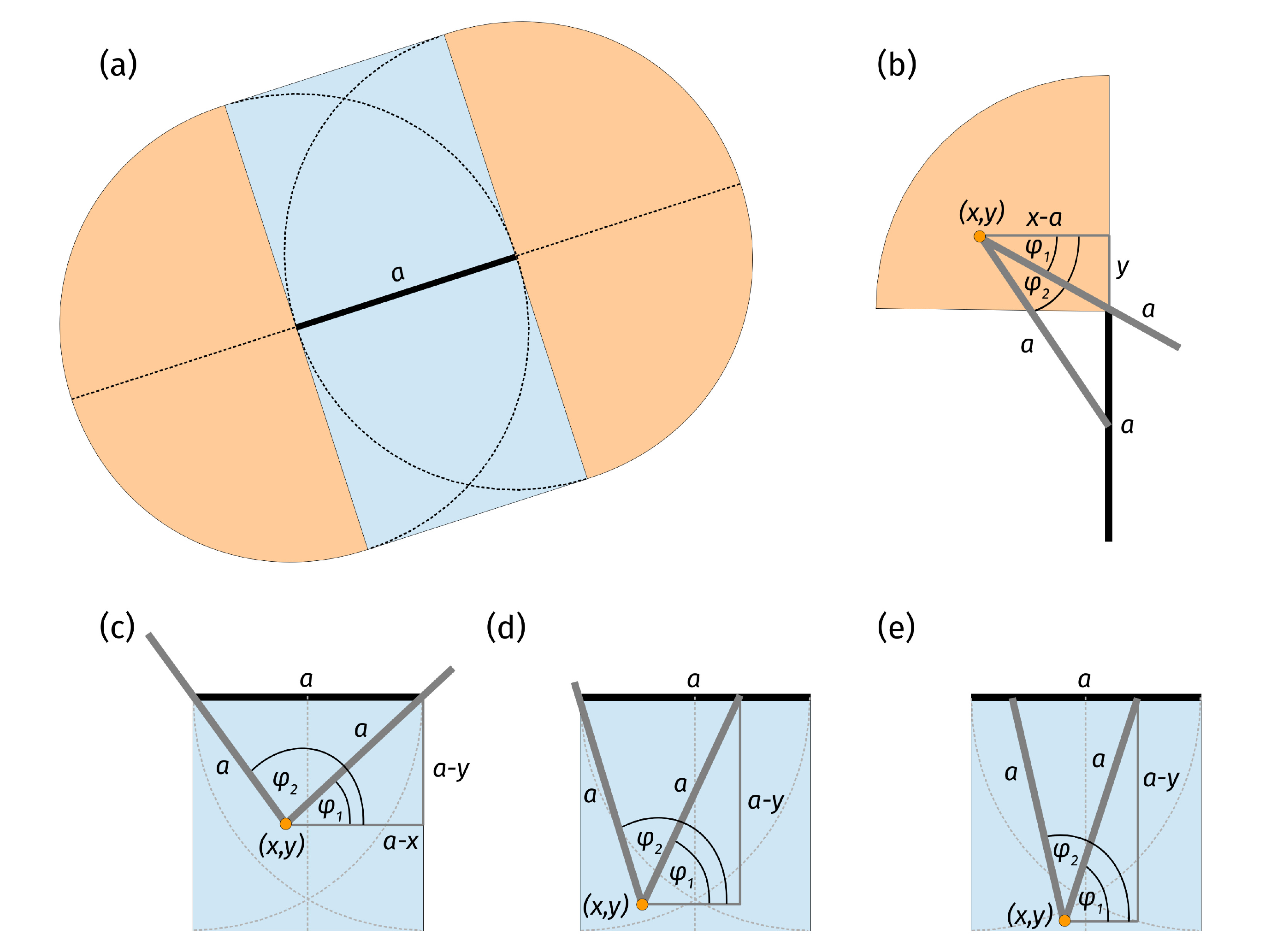}
\caption{\textbf{Intersection area.} (a) All points within the shown area have a maximal distance $a$ from the black solid line segment of length $a$. We partition the total intersection area in quarter circles (orange) and squares (blue). (b--d) The grey line segment of length $a$ within the quarter circle or square can overlap with the black line segment of length $a$ if the angle $\varphi$ lies within the interval $[\varphi_1,\varphi_2]$.}
\label{fig:overlap_area}
\end{figure}
To determine the probability that two line segments of length $a$ intersect if both line segments are uniformly at random positioned in the unit square, we divide the possible area of intersection in squares and quarter circles (see Fig.~\ref{fig:overlap_area}). In this way, we can compare the likelihoods of different geometrical intersection patterns. We first focus on the square-shaped region (see Fig.~\ref{fig:overlap_area}) that can be further sub-divided into regions I, II, and III (separated by dashed grey lines in Figs.~\ref{fig:overlap_area} (c--d)).
We consider the lower-left corner of the square-shaped region to be the origin. In region I (see Fig.~\ref{fig:overlap_area} (c)), the possible origin of an intersecting line segment is described by the coordinates $x\in[0,a/2]$ and $y\in[-\sqrt{a^2-(x-a)^2}+a,a]$. The possible intersection angles are $\varphi \in [\arctan\left(\frac{a-y}{a-x}\right),\frac{\pi}{2}+\arctan\left(\frac{x}{a-y}\right)]$ and the resulting intersection probability is
\begin{align}
\begin{split}
I_1(a)&=\frac{1}{2 \pi}\int_0^{a/2} \int_{-\sqrt{a^2-(x-a)^2}+a}^a \int_{\arctan\left(\frac{a-y}{a-x}\right)}^{\frac{\pi}{2}+\arctan\left(\frac{x}{a-y}\right)} \, \mathrm{d}x\mathrm{d} y \mathrm{d} \varphi\\
&=\frac{1}{2 \pi}\int_0^{a/2} \int_{-\sqrt{a^2-(x-a)^2}+a}^a \frac{\pi}{2}+\arctan\left(\frac{x}{a-y}\right)-\arctan\left(\frac{a-y}{a-x}\right) \, \mathrm{d}x\mathrm{d} y \mathrm{d} \varphi \\
&= \frac{1}{2 \pi} \int_0^{a/2} \int_0^{\sqrt{a^2-(x-a)^2}} \frac{\pi}{2}+\arctan\left(\frac{x}{y'}\right) \, \mathrm{d}y' \mathrm{d}x
-\int_{a/2}^{a} \int_0^{\sqrt{a^2-{x'}^2}}\arctan\left(\frac{y'}{x'}\right) \, \mathrm{d}y' \mathrm{d}x'\\
&=\frac{1}{2 \pi}\left\{\frac{1}{48} a^2 \pi (4 \pi-3 \sqrt{3})+\frac{1}{144} a^2 \left[2 \pi^2-3 \sqrt{3} \pi+18(1+\ln(2))\right] - \frac{1}{72} a^2 \left[-3 \sqrt{3} \pi + 2 \pi^2 + \ln(512)\right]\right\}\\
&=\frac{1}{144 \pi} a^2 \left( 9 - 3 \sqrt{3} \pi + 5 \pi^2\right)\,.
\end{split}
\end{align}
Next, we focus on region II (see Fig.~\ref{fig:overlap_area} (d)) where $x\in[0,a/2]$ and $y\in[-\sqrt{a^2-x^2}+a,-\sqrt{a^2-(x-a)^2}+a]$, and $\varphi \in [\arcsin\left(\frac{a-y}{a}\right),\frac{\pi}{2}+\arctan\left(\frac{x}{a-y}\right)]$. In this region, the intersection probability is
\begin{align}
\begin{split}
I_2(a)&=\frac{1}{2 \pi} \int_0^{a/2} \int_{-\sqrt{a^2-x^2}+a}^{-\sqrt{a^2-(x-a)^2}+a} \int_{\arcsin\left(\frac{a-y}{a}\right)}^{\frac{\pi}{2}+\arctan\left(\frac{x}{a-y}\right)} \, \mathrm{d}x\mathrm{d} y \mathrm{d} \varphi \\
&=\frac{1}{2 \pi} \left[\int_0^{a/2} \int_{-\sqrt{a^2-x^2}+a}^{-\sqrt{a^2-(x-a)^2}+a} \frac{\pi}{2}+\arctan\left(\frac{x}{a-y}\right)\, \mathrm{d}x\mathrm{d} y-\int_0^{a/2} \int_{-\sqrt{a^2-x^2}+a}^{-\sqrt{a^2-(x-a)^2}+a}\arcsin\left(\frac{a-y}{a}\right)\, \mathrm{d}x\mathrm{d} y\right] \\
&= \frac{1}{2 \pi} \left[\frac{1}{24} a^2 \pi (3 \sqrt{3} - \pi)+\int_0^{a/2} \int_{\sqrt{a^2-(x-a)^2}}^{\sqrt{a^2-x^2}} \arctan\left(\frac{x}{y'}\right)\, \mathrm{d} y' \mathrm{d}x - \int_0^{a/2}\int_{\sqrt{a^2-(x-a)^2}}^{\sqrt{a^2-x^2}} \arcsin\left(\frac{y'}{a}\right) \, \mathrm{d}y' \mathrm{d}x\right] \\
&= \frac{1}{2 \pi} \left[\frac{1}{24} a^2 \pi (3 \sqrt{3} - \pi)-\frac{1}{144} a^2 \left(18 - 6 \sqrt{3} \pi + \pi^2\right)-\frac{1}{144} a^2 \left(-54 + 12 \sqrt{3} \pi + \pi^2\right) \right]\\
&=\frac{1}{72 \pi} a^2 \left[3(3+\sqrt{3}\pi)-2\pi^2\right]\,.
\end{split}
\end{align}
In region III (see Fig.~\ref{fig:overlap_area} (e)), the possible origin of the intersecting line segment is given by $x\in[0,a/2]$ and $y\in[0,-\sqrt{a^2-x^2}+a]$, and possible intersection angles are $\varphi\in[\arcsin\left(\frac{a-y}{a}\right),\pi-\arcsin\left(\frac{a-y}{a}\right)]$. Thus, the intersection probability is
\begin{align}
\begin{split}
I_3(a)&=\int_0^{a/2} \int_{0}^{-\sqrt{a^2-x^2}+a} \int_{\arcsin\left(\frac{a-y}{a}\right)}^{\pi-\arcsin\left(\frac{a-y}{a}\right)} \, \mathrm{d}x\mathrm{d} y \mathrm{d} \varphi \\
&= \int_0^{a/2} \int_{0}^{-\sqrt{a^2-x^2}+a} \pi - 2 \arcsin\left(\frac{a-y}{a}\right)  \, \mathrm{d}x\mathrm{d} y\\
&=\frac{1}{144 \pi} a^2 \left[3 (9 - \sqrt{3}\pi)  - \pi^2\right]\,.
\end{split}
\end{align}
and
\begin{equation}
I_1(a)+I_2(a)+I_3(a)=\frac{3 a^2}{8 \pi} \,.
\label{eq:intersect_prob1}
\end{equation}
Finally, there are four quarter circles where $x\in[0,a]$, $y\in [0,\sqrt{a^2-(x-a)^2}]$, and $\varphi \in [\arctan\left(\frac{y}{a-x}\right),\arccos\left(\frac{a-x}{x}\right)]$ such that
\begin{equation}
\begin{split}
I_4(a)&=\frac{1}{2 \pi}\int_0^{a} \int_{0}^{\sqrt{a^2-(x-a)^2}} \int_{\arctan\left(\frac{y}{a-x}\right)}^{\arccos\left(\frac{a-x}{a}\right)} \, \mathrm{d}x\mathrm{d} y \mathrm{d} \varphi \\
&=\frac{1}{2 \pi} \left[\int_0^{a} \int_{0}^{\sqrt{a^2-(x-a)^2}} \arccos\left(\frac{a-x}{a}\right)\, \mathrm{d}x\mathrm{d} y-\int_0^{a} \int_{0}^{\sqrt{a^2-(x-a)^2}}\arctan\left(\frac{y}{a-x}\right)\, \mathrm{d}x\mathrm{d} y\right] \\
&=\frac{1}{2 \pi}\left[\frac{1}{16} a^2 (4 + \pi^2)-\frac{a^2 \pi^2}{16}\right]\\
&=\frac{a^2}{8 \pi}\,.
\end{split}
\label{eq:intersect_prob2}
\end{equation}
According to Eqs.~\eqref{eq:intersect_prob1} and \eqref{eq:intersect_prob2}, the probability that an intersection occurs from the four quarter circles is three times smaller than the intersection probability in the rectangular region (see Fig.~\ref{fig:overlap_area}).
The total intersection probability is
\begin{equation}
p(a)=4 \left[I_1(a)+I_2(a)+I_3(a)+I_4(a)\right]=\frac{2 a^2}{\pi}\,.
\end{equation}
\end{document}